\documentclass[12pt]{article}

\usepackage{amsmath}
\usepackage{amsfonts}
\usepackage{amssymb}

\newcommand{\be}{\begin{equation}}
\newcommand{\ee}{\end{equation}}
\newcommand{\ba}{\begin{eqnarray}}
\newcommand{\ea}{\end{eqnarray}}

\begin{document}

\begin{center}
{\large\bf NEW TWO-DIMENSIONAL QUANTUM MODELS WITH SHAPE INVARIANCE}\\

\vspace{0.3cm} {\large \bf F. Cannata$^{1}$\footnote{E-mail: cannata@bo.infn.it}, M.V. Iof\/fe$^{2}$\footnote{E-mail: m.ioffe@pobox.spbu.ru},
D.N. Nishnianidze$^{2,3}\footnote{E-mail: cutaisi@yahoo.com}$
}\\
\vspace{0.2cm}
$^1$INFN, Via Irnerio 46, 40126 Bologna, Italy\\
$^2$Saint-Petersburg State University,
198504 St.-Petersburg, Russia\\
$^3$Akaki Tsereteli State University, 4600 Kutaisi, Republic of Georgia\\
\end{center}

\hspace*{0.5in}

\begin{minipage}{6.0in}
{\small Two-dimensional quantum models which obey the property of shape invariance are built in the framework of
polynomial two-dimensional SUSY Quantum Mechanics. They are obtained using the expressions for known one-dimensional shape invariant
potentials. The constructed Hamiltonians are integrable with symmetry operators of fourth order in momenta, and they are not amenable to the conventional separation of variables.
\\
\vspace*{0.1cm} PACS numbers: 03.65.-w, 03.65.Fd, 11.30.Pb }
\end{minipage}

\vspace*{0.4cm}

\vspace*{0.4cm}

\setcounter{footnote}{0} \setcounter{equation}{0}
\section{Introduction.}

The method of Supersymmetric (SUSY) Quantum Mechanics \cite{cooper} provides a very effective approach for investigation
of different problems in Quantum Mechanics. The SUSY intertwining relations \cite{witten} and the notion of shape invariance
\cite{gendenstein} are
most powerful tools. In one-dimensional SUSY QM they jointly reproduced in a very elegant way
all exactly solvable models which
were known by that time \cite{sukhatme}, \cite{mallow}. In addition, the intertwining relations provide the wide class of the so-called
quasi-exactly-solvable (QES) or, equivalently, partially solvable models \cite{turbiner}. Shape invariance for models
invariant under a centrally extended superalgebra with an additional symmetry generator was demonstrated in \cite{spector}. The shape invariance
in the context of Calogero-like $N-$body systems was studied in \cite{calogero}.

In two-dimensional generalization of SUSY QM, the intertwining relations \cite{abi}, \cite{abei}, \cite{david} and shape invariance
\cite{new}, \cite{ioffe}, \cite{shape}, \cite{2shape} also play
very important role, but their realization is slightly different. First of all, it must be taken into account
that different two-dimensional generalizations of SUSY QM exist. The direct generalization \cite{abi}, \cite{abei} with first order
supercharges $Q^{\pm}$ intertwines Hamiltonians with different matrix dimensionality. We shall consider another
opportunity for the two-dimensional case - intertwining of two scalar Hamiltonians $H^{(1)},\, H^{(2)}$
by supercharges $Q^{\pm}$ of second order in derivatives \cite{david}, \cite{ioffe}:
 \ba
 &H^{(1)}(\vec{x})Q^+(\vec{x})=Q^+(\vec{x})H^{(2)}(\vec{x}),\label{1}\\
 &Q^-(\vec{x})H^{(1)}(\vec{x})=H^{(2)}(\vec{x})Q^-(\vec{x}),\label{2}\\
 &H^{(1),(2)}=-\partial_1^2-\partial^2_2 +V^{(1),(2)}(\vec x);\,\, \vec x= (x_1, x_2), \nonumber
 \ea
 where supercharges are chosen with Lorentz metrics in second order derivatives:
 \be
 Q^+=4\partial_+\partial_- + 4C_+\partial_- +4C_-\partial_+ + B(\vec{x});\, Q^-=(Q^+)^{\dagger};\,\,
 x_{\pm}=x_1\pm x_2. \label{3}
 \ee
As was shown in \cite{david}, this choice of the metrics simplifies essentially the solution of intertwining
relations (\ref{1}), (\ref{2}). In the case of arbitrary form of second order part in $Q^{\pm}$, these relations
are equivalent to the very complicated system of nonlinear
partial differential equations for the functions $V^{(1),(2)},\, C_{\pm},\, B.$ For the case of Lorentz
form (\ref{3}) of second order part, this system of equations leads to $C_{\pm}=C_{\pm}(x_{\pm})$ and to the expressions for the potentials (we remark that coefficient functions $C_{\pm}$
differ from that in previous papers \cite{david}, \cite{ioffe}, \cite{new} by a multiplier $4$):
 \ba
 &V^{(1)}(\vec{x})=2(C_+'+C_+^2+C_-'+C_-^2)+f_2(x_2)-f_1(x_1)\equiv\nonumber\\
 &\equiv v^{(1)}_+(x_+)+v^{(1)}_-(x_-)+f_2(x_2)-f_1(x_1),\label{4}\\
 &V^{(2)}(\vec{x})=2(C_+^2-C_+'+C_-^2-C_-')+f_2(x_2)-f_1(x_1)\equiv\nonumber\\&
 v^{(2)}_+(x_+)+v^{(2)}_-(x_-)+f_2(x_2)-f_1(x_1),\label{5}\\
 &B(\vec{x})=4C_+(x_+)C_-(x_-)+f_1(x_1)+f_2(x_2),\label{6}
 \ea
 where the functions $C_{\pm}(x_{\pm})$ and $F(\vec{x})\equiv
f_1(x_1)+f_2(x_2)$ have to satisfy the following equation:
 \ba
 \partial_-(C_-F)=-\partial_+(C_+F),\label{7}
 \ea
the last (unsolved yet) equation of the above mentioned system of equations.
This equation was solved in \cite{david}, \cite{ioffe} by means of choosing some simplifying ansatzes.
Some of these solutions obey \cite{shape}, \cite{2shape} two-dimensional shape invariance property.

In the present paper we shall study the general question: is there any, wide enough, additional class of shape invariant
systems? Therefore, it is timely to remind for the reader's convenience the
idea of shape invariance in SUSY Quantum Mechanics.

In one-dimensional Quantum Mechanics with first order differential operators $Q^{\pm}$,
one refers to shape invariance \cite{gendenstein}, \cite{cooper}, when the partner Hamiltonians in (\ref{1}), (\ref{2})
both depend on some (multi)parameter $a,$ and they have the similar shape, i.e. they satisfy:
\be
H^{(2)}(x, a)=H^{(1)}(x, \tilde a) + {\cal R}(a),
\label{shape1}
\ee
where $\tilde a=\tilde a(a)$ is some new value of parameter,
which depends on $a,$ and $ {\cal R}(a)$ is a ($c$-number) function of $a.$
In the case of absence of spontaneous breaking of supersymmetry, this property
allows to construct the analytical expressions for all wave functions $\Psi (x, a)$ in a pure
algebraic way (see details in \cite{cooper}). The whole energy spectrum of the Hamiltonian is found
as well. Among several relations of SUSY algebra intertwining relations (\ref{1}), (\ref{2}) play a crucial role in this approach.

The notion of shape invariance was generalized onto two-dimensional SUSY Quantum Mechanics
in \cite{new}, \cite{ioffe}, \cite{shape}, \cite{2shape}. In this case, the partner Hamiltonians depend also on (multi)parameter $a,$
the main relation (\ref{shape1}) has the same form, but operators
$Q^{\pm}$ are of second order in derivatives. Thereby, in contrast to one-dimensional situation,
$Q^{\pm}$ have many zero modes, and one has chance to find
only a part of the spectrum and the corresponding wave functions (quasi-exact-solvability) \cite{new}, \cite{ioffe}. This does not prevent us from
finding the whole spectrum (exact solvability), using however some additional tools \cite{physrev}.

The structure of the paper is the following. The general conditions which guarantee the shape invariance of two-dimensional potentials are in
Section 2, using their integrability and the results on shape invariant one-dimensional systems. The possible ansatzes for the coefficient functions of the supercharges are defined. In Section 3 the explicit expressions for potentials are derived. The particular forms of these potentials for specific values of parameters are given in Appendix. Some of potentials are recognized as already known, but others are new to the best of our knowledge.

\section{Two-dimensional shape invariance.}

The problem of constructing of most general form of shape invariant potential seems to be a rather difficult task. Even in one-dimensional SUSY Quantum Mechanics it is not yet fully solved
 (see discussion in \cite{sukhatme}, \cite{mallow}). Only a class of such potentials was built, and it coincides with variety of well known exactly solvable potentials. It was proven recently \cite{mallow} that no additional shape invariant potentials
with so called additive shape invariance exist, outside this class.

It is clear that one has not too many chances to solve the analogous problem in two-dimensional situation. Therefore, we do not pretend to find all existing
shape invariant two-dimensional potentials. At best, we can find a wide class of such new potentials. It is necessary to remind that
some shape invariant potentials were already obtained \cite{david}, \cite{ioffe} in the framework of polynomial SUSY QM. Among these potentials two-dimensional generalizations of Morse potential and P\"oschl-Teller potential must be mentioned specially \cite{new}, \cite{physrev}, \cite{valinevich}.

It is convenient, without loss of generality, to choose parameters in such a way that the shape invariance condition (\ref{shape1}) links Hamiltonians $H^{(1,2)}(\vec x; a)$
with difference (step) between parameters equal to unity, i.e.
 \be
 H^{(1)}(a+1;\vec{x})-H^{(2)}(a;\vec{x})=const.\label{8}
 \ee
It is evident from the explicit expressions (\ref{4}), (\ref{5}) for potentials, that the shape invariance (\ref{8}) is equivalent to a pair of one-dimensional shape invariance with "superpotentials" $C_{\pm}(x_{\pm}):$
\ba
v^{(1)}_{\pm}(a_{\pm}+1; x_{\pm}) - v^{(2)}_{\pm}(a_{\pm}; x_{\pm}) = c_{\pm},  \label{88}
\ea
i.e. Hamiltonians $H^{(1, 2)}$ obey actually two-parametric shape invariance with the two-component parameter $a\equiv (a_+,\,  a_-).$
Besides $v^{(1,2)}_{\pm},$ the potentials $V^{(1,2)}(a; \vec x)$ contain also the terms $f_2(x_2)-f_1(x_1),$ which prevent separation of variables.

The dependence of $H^{(1,2)}$ on $a$ seems superficially to be a generic dependence on two
independent variables $a_+,\, a_-.$ However, as will be clear from the following Eq.(\ref{17}) there is a constraint.
It is well known \cite{david}, \cite{ioffe}, that arbitrary Hamiltonians $H^{(1)}(a; \vec x), H^{(2)}(a; \vec x)$, which participate in SUSY intertwining relation, are integrable, i.e.
they obey the symmetry operators (here of fourth order in derivatives):
 \ba
 &&[R^{(2)}(a;\vec{x}),H^{(2)}(a;\vec{x})]=0,\quad R^{(2)}(a;\vec{x})=Q^+(a;\vec{x})Q^-(a;\vec{x})\nonumber\\
 &&[R^{(1)}(a;\vec{x}),H^{(1)}(a;\vec{x})]=0; \quad R^{(1)}(a;\vec{x})=Q^-(a;\vec{x})Q^+(a;\vec{x}).
\nonumber
 \ea
The shape invariance relation (\ref{8}) allows to conclude that:
 \ba
 [R^{(1)}(a+1;\vec{x})-R^{(2)}(a;\vec{x}),H^{(2)}(a;\vec{x})]=0.\label{14}
 \ea

Although the operator $R^{(1)}(a+1;\vec{x})-R^{(2)}(a;\vec{x})$ seems to be a symmetry operator for $H^{(2)}(a;\vec{x}),$
one can check that it is of second order in derivatives. The symmetry operator of second order in momenta signals that the Hamiltonian is amenable to separation of variables \cite{miller}, \cite{david}. Since from the very beginning, we do not consider systems with separation of variables,
we shall concern ourselves only with the case when this operator is a function of $H^{(2)}(a;\vec{x}),$ i.e.:
\be
 R^{(1)}(a+1;\vec{x})-R^{(2)}(a;\vec{x})=\mu H^{(2)}(a;\vec{x})+\nu,\label{15}
\ee
with $\mu, \nu$-constants.

Using expressions for $Q^{\pm},$ the l.h.s. of (\ref{15}) can be calculated explicitly:
\ba
 &&R^{(1)}(a+1)-R^{(2)}(a)=-8\biggr(v^{(1)}_+(a_++1)-v^{(2)}_+(a_+)\biggl)
 \partial_-^2-8\biggr(v^{(1)}_-(a_-+1)-v^{(2)}_-(a_-)\biggl)\partial_+^2+
 \nonumber\\
 &&+16\biggr(C'_+(a_++1)C'_-(a_-+1)-C'_+(a_+)C'_-(a_-)\biggl)+
 4\biggr(C_+(a_++1)\partial_-B(a+1)+
 \nonumber\\
 &&+C_-(a_-+1)\partial_+B(a+1)+C_+(a_+)\partial_-B(a)+
 C_-(a_-)\partial_+B(a)\biggl)+B^2(a+1)-B^2(a),\label{16}
 \ea
where the dependence on space coordinates was omitted for brevity.
Comparing with (\ref{15}), we obtain the necessary condition, which constants $c_{\pm}$ in the r.h.s of Eq.(\ref{88}) have to satisfy. Namely they should coincide:
 \be
 v^{(1)}_+(a_++1)-v^{(2)}_+(a_+)=v^{(1)}_-(a_-+1)-v^{(2)}_-(a_-)=c .\label{17}
 \ee

According to analysis of \cite{mallow}, where all known one-dimensional shape invariant potentials were reproduced, we shall consider the following form of dependence of "superpotentials" $C_{\pm}$ on parameters $a_{\pm}:$
 \be
 C_{\pm}(x_\pm)=a_{\pm}p_{\pm}(x_{\pm})+
 r_{\pm}(x_{\pm}) + q_{\pm}(a_{\pm}),\label{9}
 \ee
where $a_{\pm}$ are parameters of shape invariance with a unit step $\widetilde a_{\pm} = a_{\pm}+1,$
functions $p_{\pm}(x_{\pm}), r_{\pm}(x_{\pm})$ do not depend on $a_{\pm},$ and $q_{\pm}(a_{\pm})$ do not depend on $x_{\pm}.$
It was shown in \cite{mallow}, that it is sufficient to use the two-term variants of (\ref{9}), therefore we shall consider below separately two different kinds of possible dependence of $C_{\pm}(x_\pm)$ on shape invariance parameters:
\ba
&&I)\qquad\qquad\,\, C_{\pm}(a_{\pm})=a_{\pm}p_{\pm}+r_{\pm};
 \label{12}\\
 &&II)\qquad\qquad C_{\pm}(a_{\pm})=a_{\pm}p_{\pm}+q_{\pm}(a_{\pm}).\label{1D}
 \ea
We shall explore these ansatzes together with additional requirements, which are specific for two-dimensional systems, namely Eq.(\ref{7})
and (\ref{15}).

\section{Construction of shape invariant potentials.}

The shape invariance of first kind can be considered, starting from Eq.(\ref{17}), which for the choice (\ref{12}) reads as:
\be
(2a_{\pm}+1)(p_{\pm}^2 + p_{\pm}^{\prime})+2p_{\pm}r_{\pm}+2r_{\pm}^{\prime}=c/2, \label{pp10}
\ee
where $c$ is an arbitrary constant, the same for signs $\pm$ in the l.h.s.
It follows that in this case, $p_{\pm}(x_{\pm}),\, r_{\pm}(x_{\pm})$ satisfy the system of differential equations:
\ba
p_{\pm}^2 + p_{\pm}^{\prime} = \lambda^2_{\pm}; \quad  p_{\pm}r_{\pm}+r_{\pm}^{\prime}=d_{\pm},
\nonumber
\ea
and its general solution is:
\ba
p_{\pm}&=&\frac{z^{\prime}_{\pm}(x_{\pm})}{z_{\pm}(x_{\pm})};\quad z_{\pm}(x_{\pm})=\sigma_{\pm}\exp{(\lambda_{\pm}x_{\pm})} +
\delta_{\pm}\exp{(-\lambda_{\pm}x_{\pm})}; \label{pp12}\\
r_{\pm}&=&\frac{1}{z_{\pm}(x_{\pm})}\biggl( \alpha_{\pm} + d_{\pm}\int^{x_{\pm}}z_{\pm}(x_{\pm})dx_{\pm} \biggr)=
\frac{1}{z_{\pm}(x_{\pm})}\biggl( \alpha_{\pm} + \frac{d_{\pm}}{\lambda_{\pm}^2}z^{\prime}_{\pm}(x_{\pm}) \biggr), \label{pp13}
\ea
with integration constants $\alpha_{\pm}.$ The form (\ref{12}) of $C_{\pm}$ and expressions (\ref{pp12}), (\ref{pp13}) in terms of
 $z,\, z^{\prime}$ give an opportunity to take $d_{\pm}=0$ by means of transformation of parameter $a_{\pm}$ to $(a_{\pm}+d_{\pm}/\lambda^2_{\pm}).$
 Thus, we shall continue with $d_{\pm}=0$ below. Also, the direct calculations give that the constant $c$ in the r.h.s. of Eq.(\ref{17}) is:
 $c=2(2a+1)\lambda_{\pm}^2,$ and therefore $\lambda_+=\lambda_-\equiv\lambda .$

Now we can go to the system of equations (\ref{7}). Since the function $F$ does not
depend on $a_{\pm},$ it follows from (\ref{7}) for independent parameters $a_+$ and $a_-,$ that the following relations must be
fulfilled:
\be
\partial_{+}(F p_{+})= \partial_{-}(F p_{-})= 0;\quad \partial_{+}(F r_+)+\partial_{-}(F r_-)=0.
\label{10}
\ee
From the first two equalities, one can conclude that $F=Const /(p_+(x_+) p_-(x_-)).$  This is just a case of factorizable function $F,$ which was
studied earlier in \cite{david} in a general form. One of two solutions, which were found there, indeed obeys shape invariance, but it is amenable to standard separation of variables. By this reason, we shall not consider this case further, restricting ourselves to a case with one independent parameter instead of two: $a_-=a_+\equiv a.$ As a consequence, the possible shape invariance is still reduced to a pair of one-dimensional shape invariance in variables $x_{\pm},$ but the restrictions onto functions become much less strong than (\ref{10}) now:
\ba
&&\partial_+(F p_+)+\partial_-(F p_-)=0;  \label{34}\\
&&\partial_+(Fr_+)+\partial_-(Fr_-)=0. \label{35}
\ea

If one of the constants $\alpha_{+}$ or $\alpha_-$ in (\ref{pp13}) vanishes, it is clear from (\ref{35}) that $F$ is factorizable, therefore
we shall study two other choices of constants:
$$\quad Ia)\,\,\quad\alpha_-=\alpha_+=0;\qquad Ib)\,\,\quad \alpha_-\alpha_+\neq 0.$$

{\bf Ia).} For this option, Eq.(\ref{35}) is satisfied identically, but we have no direct way to solve Eq.(\ref{34})
in a general form. We shall act in an indirect way. It follows from (\ref{16}), that
\ba
 &&R^{(1)}(a+1)-R^{(2)}(a)=8\lambda^2(2a+1)\biggr(H(a)+2\lambda^2(2a+1)\biggl)+
 \nonumber\\
 &&+2(2a+1)\biggr(4\lambda^2(f_1-f_2)+(p_++p_-)f'_1+(p_--p_+)f'_2+4p_+p_-(f_1+f_2)
 \biggl)+
 \nonumber\\
 &&+4\biggr((r_++r_-)f'_1+(r_--r_+)f'_2+
 2(p_+r_-+p_-r_+)(f_1+f_2)\biggl).\label{36}
\ea
According to Eq.(\ref{15}) and since the functions $f_{1,2}, p_{\pm}$
do not depend on parameter $a,$ each of last two terms in (\ref{36}) must be constant. In particular,
\be
 4\lambda^2(f_1-f_2)+(p_++p_-)f'_1+(p_--p_+)f'_2+4p_+p_-(f_1+f_2)\equiv 2\omega;\,\, \omega = const.
\nonumber
\ee
Together with (\ref{34}), this equation gives:
\be
 f_2(x_2)=\frac{4\lambda^2f_1(x_1)-\omega + (p_++p_-)f'_1(x_1)}{(p_+-p_-)^2}-f_1(x_1),\label{38}
\ee
and after differentiation over $x_1,$ we obtain the equation for the function $f_1(x_1):$
 \ba
 f''_1+\frac{6(\lambda^2+p_+p_-)}{p_++p_-}f'_1+8\lambda^2f_1=2\omega.\label{39}
 \ea
From definition (\ref{pp12}) of $p_{\pm}:$
\be
 \frac{6(\lambda^2+p_+p_-)}{p_++p_-}=\frac{3z'_1}{z_1},\label{40}
\ee
where $z_1$ is defined as:
\be
z_1(x_1)\equiv \sigma_+\sigma_-\exp{(2\lambda x_1)} - \delta_+\delta_-\exp{(-2\lambda x_1)};\quad z_1^{\prime\prime}=
4\lambda^2z_1. \label{z1z1}
\ee
Eq.(\ref{40}) transforms (\ref{39}) as follows:
\be
 (f''_1z_1+z'_1f'_1)+2(z'_1f'_1+z''_1f_1)=\frac{\omega}{2\lambda^2}z''_1,
\nonumber
\ee
whose solution is:
\be
 f_1(x_1)=\frac{\omega}{4\lambda^2}+\frac{k_1z'_1+k_2}{z_1^2}, \label{42}
\ee
with $k_1,\,k_2 -$ constants.
Substitution back into (\ref{38}) gives:
\be
 f_2(x_2)=-\frac{\omega}{4\lambda^2}+\frac{k_1z'_2-k_2}{z_2^2},\label{43}
\ee
with the definition of $z_2:$
\be
z_2(x_2)\equiv \sigma_+\delta_-\exp{(2\lambda x_2)} - \delta_+\sigma_-\exp{(-2\lambda x_2)};\quad z_2^{\prime\prime}=
4\lambda^2z_2. \label{z2z2}
\ee
It follows from (\ref{42}) and (\ref{43}), that we may take the value $\omega =0,$ since $f_1$ and $f_2$
are defined up to an additive constant with opposite sign.

Thus, the option Ia) gives the following shape invariant potentials:
\ba
 &&V^{(1),(2)}(a;\vec{x})=-8a(a\mp 1)\lambda^2\biggr(\frac{\sigma_+\delta_+}
 {(\sigma_+\exp(\lambda x_+)+\delta_+\exp(-\lambda
 x_+))^2}+
 \nonumber\\
 &&+\frac{\sigma_-\delta_-}{(\sigma_-\exp(\lambda x_-)+\delta_-\exp(-\lambda
 x_-))^2}\biggl)-
 \frac{k_1(\sigma_+\sigma_-\exp(2\lambda x_1)+
 \delta_+\delta_-\exp(-2\lambda x_1))+k_2}
 {(\sigma_+\sigma_-\exp(2\lambda
 x_1)-\delta_+\delta_-\exp(-2\lambda x_1))^2}+\nonumber\\
 &&+\frac{k_1(\sigma_+\delta_-\exp(2\lambda
 x_2)+\delta_+\sigma_-\exp(-2\lambda x_2))-k_2}
 {(\sigma_+\delta_-\exp(2\lambda
 x_2)-\delta_+\sigma_-\exp(-2\lambda x_2))^2}.\label{44}
 \ea
The particularized expressions for various choices of the arbitrary constants are given in the Appendix.

{\bf Ib).} In this case we shall solve the system of equations (\ref{34}), (\ref{35})
directly. This task is simplified by assuming $\alpha_+=\alpha_-\equiv\alpha $ in (\ref{pp13}) without loss of
generality. This is possible due to homogeneity of equations (\ref{34}), (\ref{35}) under multiplication of $r_{\pm}$ by constant
factor. After simple manipulations we obtain the system (\ref{34}), (\ref{35}) in the form:
\ba
 \partial_1(\ln F(p_+r_--r_+p_-))=\frac{r'_-p_--p'_-r_-+r_+p'_+-r'_+p_+}{p_+r_--r_+p_-}
 =\frac{\lambda^2(r_+-r_-)}{p_+r_--r_+p_-},
\nonumber\\
 \partial_2(\ln F(p_+r_--r_+p_-))=\frac{r'_-p_--p'_-r_--r_+p'_++r'_+p_+}{p_+r_--r_+p_-}
 =-\frac{\lambda^2(r_++r_-)}{p_+r_--r_+p_-},
\nonumber
\ea
 and using expressions (\ref{pp12}), (\ref{pp13}):
 \ba
 \partial_1(\ln F(p_+r_--r_+p_-))=-\lambda^2\frac{y_+-y_-}{y'_+-y'_-}=
 -\frac{y''_+-y''_-}{y'_+-y'_-}=
 -\partial_1\ln(y'_+-y'_-).
\nonumber\\
 \partial_2(\ln F(p_+r_--r_+p_-))=-\lambda^2\frac{y_++y_-}{y'_+-y'_-}=
 -\frac{y''_++y''_-}{y'_+-y'_-}=
 -\partial_2\ln (y'_+-y'_-).
\nonumber
 \ea
These equations can be integrated explicitly:
 \ba
 F=\frac{y_+y_-}{(y'_+-y'_-)^2},\label{52}
 \ea
but it is necessary to take into account additionally that $F=f_1(x_1)+f_2(x_2).$ This gives
restriction onto parameters in (\ref{pp12}): $\sigma_+\delta_+=\sigma_-\delta_-.$
Then, for $\delta_+\neq 0,$ we obtain from (\ref{52}). that:
 \ba
 f_1=\frac{4k\sigma_-}{(\sigma_-\exp(\lambda x_1)+\delta_+\exp(-\lambda
 x_1))^2},
\nonumber\\
 f_2=\frac{4k\delta_-}{(\delta_-\exp(\lambda x_2)-\delta_+\exp(-\lambda
 x_2))^2},
\nonumber
 \ea
 and corresponding shape invariant potentials are:
 \ba
 &&V^{(1),(2)}(a;\vec{x})=-4\biggl(2\lambda^2a(a\mp 1)\sigma_-\delta_- -\alpha^2\biggr)\cdot\nonumber\\
&&\cdot\biggl(\frac{\delta_+^2}{(\sigma_-\delta_-\exp(\lambda x_+)+
 \delta_+^2\exp(-\lambda x_+))^2}+
\frac{1}{(\sigma_-\exp(\lambda x_-)+\delta_-\exp(-\lambda
 x_-))^2}\biggr)\nonumber\\
 &&+4\alpha(2a\mp 1)\lambda
 \biggr(\frac{\delta_+((\sigma_-\delta_-\exp(\lambda
 x_+)-\delta_+^2\exp(-\lambda x_+))}{(\sigma_-\delta_-\exp(\lambda x_+)+
 \delta_+^2\exp(-\lambda x_+))^2}+
 \nonumber\\
 &&+\frac{\sigma_-\exp(\lambda x_-)-\delta_-\exp(-\lambda
 x_-)}{(\sigma_-\exp(\lambda x_-)+\delta_-\exp(-\lambda
 x_-))^2}\biggl)+4k\biggr(\frac{\delta_-}
 {(\delta_-\exp(\lambda x_2)-\delta_+\exp(-\lambda
 x_2))^2}-
 \nonumber\\
 &&-\frac{\sigma_-}{(\sigma_-\exp(\lambda x_1)+\delta_+\exp(-\lambda
 x_1))^2}\biggl).\label{55}
 \ea
 Explicit expressions for these potentials for particular choices of the arbitrary constant parameters
 are presented in the Appendix.
 By the way, Eq.(\ref{16}) for potentials (\ref{44}) and (\ref{55}) has the form:
\ba
 R^{(1)}(a+1)-R^{(2)}(a)=8\lambda^2(2a+1)\biggr(H^{(2)}(a)+
 2\lambda^2(2a^2+2a+1)\biggl).
\nonumber
\ea

As for the second shape invariance (\ref{1D}), Eq.(\ref{17}) allows the general solution:
\ba
&&p_{\pm}^2+p_{\pm}^{\prime}=\lambda_{\pm}^2;\quad p_{\pm}=\frac{z_{\pm}^{\prime}}{z_{\pm}};\quad z_{\pm}^{\prime\prime}=\lambda_{\pm}^2z_{\pm};
\nonumber\\
&&(a+1)q_{\pm}(a+1)+a q_{\pm}(a)=0;\quad q_{\pm}(a) = \frac{\gamma_{\pm}}{a};
\nonumber
\ea
with restriction for constants:
\be
a^2(a+1)^2\lambda_+^2-\gamma_+^2=a^2(a+1)^2\lambda_-^2-\gamma_-^2.
\nonumber
\ee
In turn, Eq.(\ref{7}) gives:
 \ba
 &&\partial_+(F p_+)+\partial_-(F p_-)=0; \label{4D}\\
 && \gamma_+\partial_+F+\gamma_-\partial_-F=0;\quad  (\gamma_++\gamma_-)f_1'(x_1)+(\gamma_+-\gamma_-)f_2'(x_2)=0.\label{6D}
 \ea
It is easy to check, that for $\gamma_+\neq \pm\gamma_-$ Eq.(\ref{6D}) gives $f_{1,2}(x_{1,2})\sim x_{1,2},$
and Eq.(\ref{4D}) can not be satisfied.

Thus, the sole opportunity $\gamma_+=\pm\gamma_-$ remains open. Then, either $f_1(x_1)$ or $f_2(x_2)$ vanishes, and $\lambda_+^2=\lambda_-^2\equiv
\lambda^2.$ After all, Eq.(\ref{4D}) can not be fulfilled, and no shape invariant potentials correspond to $C_{\pm}$ of the form (\ref{1D}).

\section{Acknowledgements.}

M.V.I. and D.N.N. are indebted to INFN, the University of
Bologna for the support and hospitality. The work of M.V.I. was partially supported by the RFFI grant 09-01-00145-a.

\section{Appendix: Potentials with specific parameters\\ values.}

Here we give the explicit formulas of shape invariant potentials for specific values of the parameters.

The general form (\ref{44}) provides the following expressions for different choices of constants:

 \ba
 &&\sigma_+=0,\delta_+=\sigma_-=-\delta_-=1,\nonumber\\
 &&V^{(1),(2)}=\frac{2\lambda^2a(a\mp 1)}{\sinh^2(\lambda x_-)}+
 k_1\biggl(\exp(2\lambda x_1)+\exp(2\lambda x_2)\biggr)-k_2\biggl(\exp(4\lambda x_1)+\exp(4\lambda x_2)\biggr).\label{44.1}\\
 &&\nonumber\\
 &&\sigma_+=0,\delta_+=\sigma_-=\delta_-=1,\nonumber\\
 &&V^{(1),(2)}=-\frac{2\lambda^2a(a\mp 1)}{\cosh^2(\lambda x_-)}-
 k_1\biggl(\exp(2\lambda x_1)-\exp(2\lambda x_2)\biggr)-\nonumber\\
&&-k_2\biggl(\exp(4\lambda x_1)+\exp(4\lambda x_2)\biggr);\label{44.2}\\
&&\nonumber\\
 &&\sigma_+=\delta_+=\sigma_-=\delta_-=1,\nonumber\\
 &&V^{(1),(2)}=-2\lambda^2a(a\mp 1)\biggl(\frac{1}{\cosh^2(\lambda
 x_-)}+\frac{1}{\cosh^2(\lambda x_+)}\biggr)-\nonumber\\
 &&-\frac{2k_1+k_2}{16}
 \biggl(\frac{1}{\sinh^2(\lambda
 x_1)}-\frac{1}{\cosh^2(\lambda x_2)}\biggr)-\frac{2k_1-k_2}{16}
 \biggl(\frac{1}{\cosh^2(\lambda
 x_1)}-\frac{1}{\sinh^2(\lambda x_2)}\biggr);\label{44.3}\\
&&\nonumber\\
 &&\sigma_+=-\delta_+=\sigma_-=-\delta_-=1,\nonumber\\
 &&V^{(1),(2)}=2\lambda^2a(a\mp 1)\biggl(\frac{1}{\sinh^2(\lambda
 x_-)}+\frac{1}{\sinh^2(\lambda x_+)}\biggr)-\nonumber\\
 &&-\frac{2k_1+k_2}{16}
 \biggl(\frac{1}{\sinh^2(\lambda
 x_1)}+\frac{1}{\sinh^2(\lambda x_2)}\biggr)-\frac{2k_1-k_2}{16}
 \biggl(\frac{1}{\cosh^2(\lambda
 x_1)}+\frac{1}{\cosh^2(\lambda x_2)}\biggr);\label{44.4}\\
&&\nonumber\\
 &&\sigma_+=\delta_+=\sigma_-=-\delta_-=1,\nonumber\\
 &&V^{(1),(2)}=-2\lambda^2a(a\mp 1)\biggl(\frac{1}{\cosh^2(\lambda
 x_+)}-\frac{1}{\sinh^2(\lambda x_-)}\biggr)-
\frac{2k_1\sinh(2\lambda x_1)+k_2}{4\cosh^2(2\lambda x_1)}-\nonumber\\
&&- \frac{2k_1\sinh(2\lambda x_2)+k_2}{4\cosh^2(2\lambda x_2)}.\label{44.5}
\ea

The analogous family of potentials for the general form (\ref{55}) is:
 \ba
 &&\sigma_-=0, \quad \delta_+=\delta_-=1,\nonumber\\
 &&V^{(1),(2)}=-4\alpha\lambda(2a\mp 1)\biggl(\exp(\lambda x_+)+\exp(\lambda
 x_-)\biggr)+\nonumber\\
 &&+4\alpha^2\biggl(\exp(2\lambda x_+)+\exp(2\lambda
 x_-)\biggr)+\frac{k}{\sinh^2(\lambda x_2)};\label{55.1}\\
&&\nonumber\\
 &&\sigma_-=0, \quad \delta_+=-\delta_-=1,\nonumber\\
 &&V^{(1),(2)}=-4\alpha\lambda(2a\mp 1)\biggr(\exp(\lambda x_+)-\exp(\lambda
 x_-)\biggl)+\nonumber\\
 &&+4\alpha^2\biggr(\exp(2\lambda x_+)+\exp(2\lambda
 x_-)\biggl)+\frac{k}{\cosh^2(\lambda x_2)};\label{55.2}\\
&&\nonumber\\
&&\sigma_-=\delta_-=\delta_+=1,\nonumber\\
&&V^{(1),(2)}=k\biggr(\frac{1}{\sinh^2(\lambda x_2)}-
\frac{1}{\cosh^2(\lambda x_1)}\biggl)+\frac{2\alpha\lambda(2a\mp 1)
\sinh(\lambda x_+)-2\lambda^2a(a\mp 1)+\alpha^2}{\cosh^2(\lambda x_+)}+
\nonumber\\&&+
\frac{2\alpha\lambda(2a\mp 1)
\sinh(\lambda x_-)-2\lambda^2a(a\mp 1)+\alpha^2}{\cosh^2(\lambda
x_-)}\label{55.3}\\
&&\nonumber\\
 &&\sigma_-=-\delta_-=\delta_+=1,\nonumber\\
 &&V^{(1),(2)}=\frac{2\lambda^2a(a\mp 1)-2\alpha\lambda(2a\mp 1)\cosh(\lambda x_+)+\alpha^2}
 {\sinh^2(\lambda x_+)}+\nonumber\\
&&+ \frac{2\lambda^2a(a\mp 1)+ 2\alpha\lambda(2a\mp 1)\cosh(\lambda x_-)+\alpha^2}
 {\sinh^2(\lambda x_-)}-
 \nonumber\\
&&- k\biggl(\frac{1}{\cosh^2(\lambda x_1)}+
 \frac{1}{\cosh^2(\lambda x_2)}\biggr);\label{55.4}\\
&&\nonumber\\
 &&\sigma_-=-\delta_-=-\delta_+=1,\nonumber\\
 &&V^{(1),(2)}=\frac{2\lambda^2a(a\mp 1)+2\alpha\lambda (2a\mp 1)\cosh(\lambda x_+)+\alpha^2}
{\sinh^2(\lambda x_+)}+\nonumber\\
&& +\frac{2\lambda^2a(a\mp 1)+2\alpha\lambda (2a\mp 1)\cosh(\lambda x_-)+\alpha^2}{\sinh^2(\lambda x_-)}-
 \nonumber\\
&&- k\biggl( \frac{1}{\sinh^2(\lambda x_2)}+
\frac{1}{\sinh^2(\lambda x_1)}\biggr).
\label{55.5}
\ea

Some of these potentials are already known. In particular, Eq.(\ref{44.1}) reproduces the two-dimensional generalization
of Morse potential. It was investigated with essential use of its shape invariance in the framework of polynomial two-dimensional SUSY Quantum Mechanics in \cite{new}, \cite{ioffe}, \cite{physrev}. The same conclusion concerns the potential (\ref{55.1}) after substitution
$x_+\equiv y_1,\, x_-\equiv y_2.$

The analogous SUSY approach was used in \cite{valinevich} for the study
of generalized two-dimensional P\"oschl-Teller potential, which coincides with (\ref{44.3}) after a suitable change of constants.

Both potential (\ref{44.4}) and (after replacement $x_-\equiv y_1,\, x_+\equiv y_2$) potential (\ref{55.5}) are the particular
cases of $N-$particle models of $BC_N$ type investigated in \cite{perelomov} in a different approach beyond the shape invariance context.

To the best of our knowledge, all other cases from the above list - the potentials (\ref{44.2}), (\ref{44.5}), (\ref{55.2}), (\ref{55.3}) and (\ref{55.4}) - are new.
An additional remark is appropriate here. The transformations of coordinates $x_1, x_2$ to $y_1=x_-, y_2=x_+$ and back link potentials (\ref{44.2})
with (\ref{55.2}) and potentials (\ref{44.5}) with (\ref{55.3}). However the coupling constants in the equivalent potentials are different, leading to existence of double shape invariance for them, similarly to that studied in \cite{2shape} for the Morse potential.

\end{document}